\documentclass[twocolumn,pre,twoside,superscriptaddress,floatfix]{revtex4-1}

\usepackage{amsmath}
\usepackage{amssymb}
\usepackage{graphicx}
\usepackage{dcolumn}
\usepackage{bm}
\usepackage{color}
\usepackage{multirow}
\usepackage{hyperref}
\usepackage{epstopdf}

\usepackage[utf8]{inputenc}
\usepackage[T1]{fontenc}
\usepackage{microtype}
\usepackage{txfonts}

\DeclareMathOperator\erf{erf}

\definecolor{orange}{rgb}{1,0.5,0}

\begin{document}


\title{Electrostatic interaction of particles trapped at fluid interfaces:\\Effects of geometry and wetting properties}

\author{Arghya Majee}
\email{majee@is.mpg.de}
\affiliation{Max-Planck-Institut f\"ur Intelligente Systeme, Heisenbergstr.\ 3, 70569 Stuttgart, Germany}
\affiliation{IV. Institut f\"{u}r Theoretische Physik, Universit\"{a}t Stuttgart, Pfaffenwaldring 57, 70569 Stuttgart, 
             Germany}
\author{Markus Bier}
\email{bier@is.mpg.de}
\affiliation{Max-Planck-Institut f\"ur Intelligente Systeme, Heisenbergstr.\ 3, 70569 Stuttgart, Germany}
\affiliation{IV. Institut f\"{u}r Theoretische Physik, Universit\"{a}t Stuttgart, Pfaffenwaldring 57, 70569 Stuttgart, 
             Germany}
\affiliation{Fakult\"at Angewandte Natur- und Geisteswissenschaften, Hochschule f\"ur angewandte Wissenschaften 
             W\"urzburg-Schweinfurt, Ignaz-Sch\"on-Str.\ 11, 97421 Schweinfurt, Germany}
\author{S. Dietrich}
\affiliation{Max-Planck-Institut f\"ur Intelligente Systeme, Heisenbergstr.\ 3, 70569 Stuttgart, Germany}
\affiliation{IV. Institut f\"{u}r Theoretische Physik, Universit\"{a}t Stuttgart, Pfaffenwaldring 57, 70569 Stuttgart, 
             Germany}

\date{29 November 2018}

\begin{abstract}
The electrostatic interaction between pairs of spherical or macroscopically long, parallel cylindrical 
colloids trapped at fluid interfaces is studied theoretically for the case of small inter-particle separations. 
Starting from the effective interaction between two planar walls and 
by using the Derjaguin approximation, we address the issue of how the electrostatic 
interaction between such particles is influenced by their curvatures and by the wetting contact angle at their surfaces.  
Regarding the influence of curvature, our findings suggest that the discrepancies between linear and nonlinear 
Poisson-Boltzmann theory, which have been noticed before for planar walls, also occur for spheres and 
macroscopically long, parallel cylinders, though their magnitude depends on the wetting contact angle. 
Concerning the influence of the wetting contact angle $\theta$ simple relations are obtained for equally 
sized particles which indicate that the inter-particle force varies significantly with $\theta$ only within 
an interval around $90^\circ$. This interval depends on the Debye length of the fluids and on the size of 
the particles but not on their shape. For unequally sized particles, a more complicated relation is obtained 
for the variation of the inter-particle force with the wetting contact angle.
\end{abstract}

\maketitle


\section{Introduction}

Colloidal particles trapped at a fluid interface usually adopt configurations which are 
energetically more favorable compared to those occurring in the adjacent bulk phase(s) \cite{Ram03, Bin06}. 
This can be exploited for a wide spectrum of systems ranging from micrometer down to nanometer 
in size and from biological to industrial processes, including the stabilization of Pickering 
emulsions \cite{Pic07}, the transport of drugs and nutrients in biological systems \cite{Din02}, the 
formation of artificial cells \cite{Li13}, oil recovery, water purification, mineral processing, 
maintaining proper foaminess of cosmetic and food products \cite{Bin06}, and the fabrication of
various nanostructured devices \cite{Bok07, Rey16}. The trapping phenomenon depends on the wetting 
properties, the size, and the shape of the colloidal particles, because it hinges on the particle-mediated 
reduction of the fluid-fluid interfacial area, and consequently on the net reduction of the free 
energy of the system \cite{Bin06}.

\begin{figure}[!t]
\begin{center}
\includegraphics[width=8.0cm]{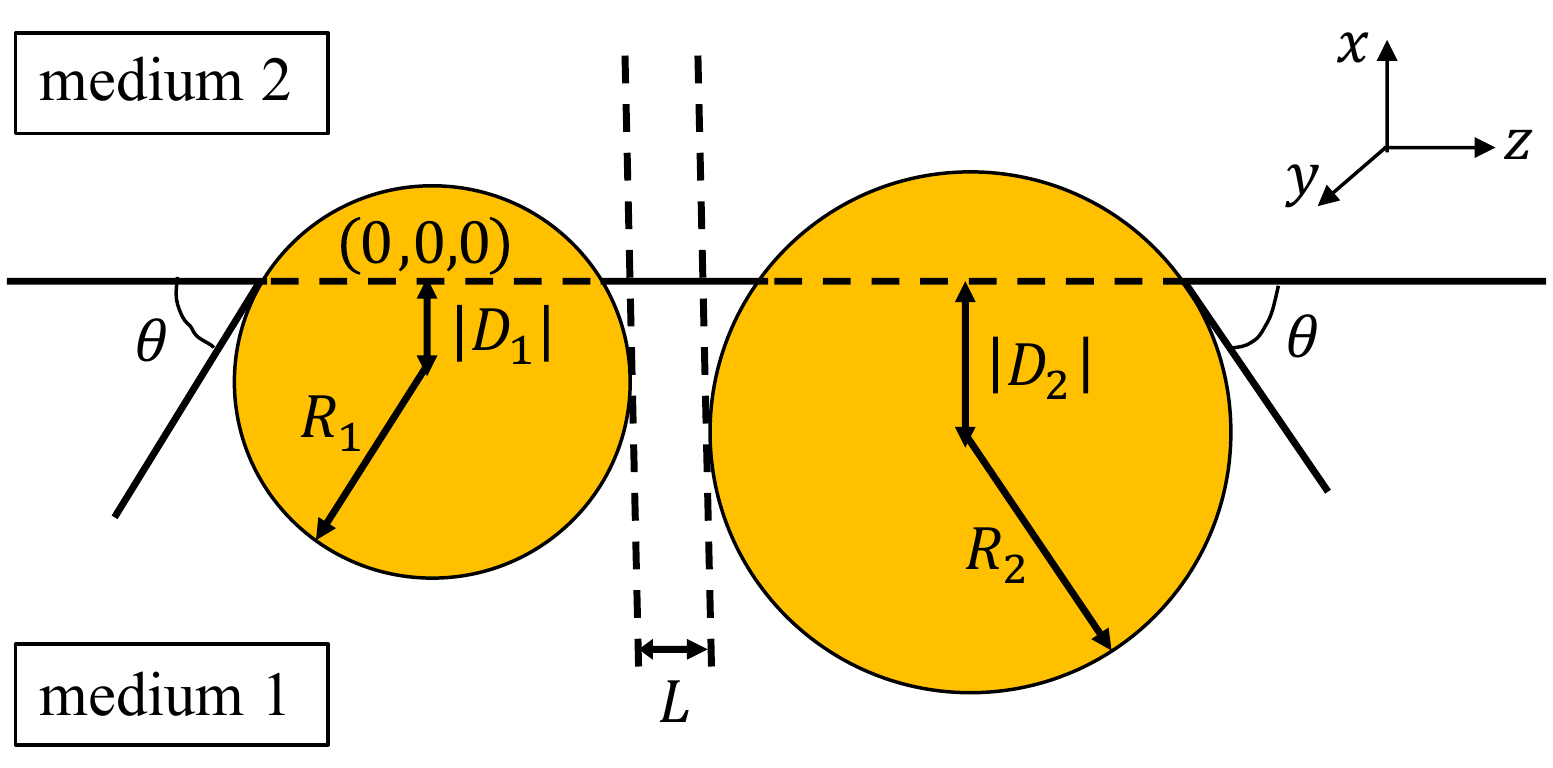}
\caption{Cross-sectional view in the $y=0$ plane of a system with two spherical or parallel 
         cylindrical colloids (yellow discs) floating at a fluid interface indicated by the 
         horizontal line. The projection of the center of the left particle 
         onto the interface is chosen as the center $(0,0,0)$ of the Cartesian coordinate system used to 
         describe the system. The fluid medium below the interface is called medium ``1'' and 
         the one above the interface medium ``2''. The particles are chemically identical, 
         such that the contact surfaces with one of the media carry the same charge densities 
         and the same wetting contact angles $\theta$. The particles may differ in size 
         with radius $R_1$ on the left and $R_2$ on the right particle. The horizontal 
         distance of the particles is characterized by the width $L$ of the gap between 
         both particles as depicted by the vertical dashed lines. 
         The equilibrium heights of the centers of the left and right particles 
         from the interface are given by $|D_1|$ and $|D_2|$, respectively, with $\theta\in\left[0,\pi\right]$ 
         determining the sign of $D_1$ and $D_2$ according to the relations $D_1=-R_1\cos\theta$ 
         and $D_2=-R_2\cos\theta$.}
\label{Fig1}
\end{center}
\end{figure}

On a mesoscopic level, the wetting properties of a colloidal particle are described best by the 
contact angle $\theta$ of the fluid-fluid interface with respect to the particle surface (see 
Fig.~\ref{Fig1}). Following the standard convention, we measure $\theta$ inside the more polar 
phase. For oil-water systems this implies that a particle is hydrophilic if $\theta<90^\circ$, 
hydrophobic if $\theta>90^\circ$, and neutrally wetted for $\theta=90^\circ$. Within the continuum 
model, the equilibrium contact angle $\theta$ of a particle is determined solely by energies 
associated with the three interfaces (two particle-liquid and one liquid-liquid) according to the 
well-known Young equation \cite{You05, Gen85}. In general, particles, which are partially wetted 
by both fluid phases, attach most stably to an interface because the corresponding trapping 
energy often exceeds several orders of $k_BT$ \cite{Bin06, Ave03}. 

Wetting properties are crucial not only for the adsorption of a single particle at an interface, but also 
for the interaction between several of them. For example, whereas very hydrophobic silica particles 
($\theta\geq129^\circ$) form well-ordered monolayer structures (with inter-particle separations of several 
particle diameter) at octane-water interfaces, less hydrophobic ($\theta\leq115^\circ$) particles fail to do 
so \cite{Hor03, Hor05}. This can be attributed to different strengths of the repulsive electrostatic force, 
which acts mainly through the oil-phase because the electrostatic field is well screened inside the aqueous 
phase at the high salt concentrations used. Moreover, the capillary interaction due to the overlap of the 
interface deformation field around each particle also depends on the contact angle $\theta$. However, 
here we disregard deformations of the interface, which can be significant if the particle surfaces are rough \cite{Sta00} 
or the particles are large (radius $\gtrsim10\,\mathrm{\mu m}$) \cite{Kra00, Oet05, Lou05, Oet08, Par15, Ana16}, 
and we focus only on the electrostatic interaction between the colloids.

Whereas the electrostatic interaction between particles trapped at a fluid interface has been studied 
extensively since the pioneering studies by Pieranski \cite{Pie80} and Hurd \cite{Hur85}, most of the 
investigations deal with the case of particles situated far away from each other. At long distances the 
electrostatic pair-interaction takes the particularly simple form of an interaction between two electric 
dipoles, which are generated by the asymmetric counterion distribution at the particle surfaces in contact 
with the two fluid phases. It has been shown that in this case the linearization of the Poisson-Boltzmann (PB) 
theory is applicable \cite{Hur85}. Recent studies, directed towards the opposite limit 
of small inter-particle separations, have been performed within the linearized PB 
theory \cite{Maj14, Lia16, Sch18} or by considering a flat plate geometry \cite{Maj14, Maj16, Sch18} in order to simplify 
the problem. Whereas the former approximation is often violated at short 
separations, the latter represents the ideal situation of a contact angle of exactly $90^\circ$ and the 
absence of particle curvature. However, in reality, the contact angle $\theta$ can vary significantly to 
either side of $90^\circ$ \cite{Ave03, Bin06, Boo15, Ana16} and whether the particle curvature plays any 
important role at short inter-particle separations still remains to be addressed within 
the nonlinear PB theory. 

Accordingly, in this contribution, we investigate the electrostatic interaction between spherical and 
parallel cylindrical colloids with an arbitrary contact angle appearing at a fluid interface. As 
long as the size of the particles is sufficiently larger than both the length scale of the interaction 
and the inter-particle separation, which is usually the case for short inter-particle 
separations we are interested in, one can apply the Derjaguin approximation (DA) using results of 
the corresponding case of planar walls \cite{Rus89}. Having recently solved this two-plate problem 
\textit{exactly} (i.e., without using the superposition approximation) within the \textit{nonlinear} 
PB theory \cite{Maj16}, we proceed one step further and compute the force between a pair of spheres 
or parallel cylinders. A similar approach has already been employed and proved to be valid in this 
context \cite{Lyn92, Lop00}. However, the present study differs from those treatments in several 
aspects. Whereas Ref. \cite{Lyn92} deals with macroscopically long cylinders trapped at an oil-water 
interface and having a constant surface potential, which is most suitable for metallic particles, we 
consider dielectric particles described by constant charge densities at their surfaces. Moreover, to 
keep our analysis general and to be consistent with experimental 
observations \cite{Ave00, Ave02, Par08, Gao14, Kel15}, we consider the particle 
surfaces to be charged in both fluid phases, which is not the case in Ref.~\cite{Lyn92}. On the other 
hand, Ref.~\cite{Lop00} describes the interaction between spherical particles by using the superposition 
approximation, which has been shown to be qualitatively wrong for small inter-particle 
separations, \cite{Maj14} and it discards any interaction between the particle-water 
surface of one particle and the particle-oil surface of the other particle. As explained in the next 
section, this latter contribution to the interaction energy is included in our calculation via 
the line contribution. 

\section{Model and formalism}

As depicted in Fig.~\ref{Fig1}, we consider two particles with radii $R_1$ and $R_2$ placed at a 
fluid-fluid interface described by a three-dimensional Cartesian coordinate system. The projection 
of the center of the left particle onto the interface is chosen as the origin $(0,0,0)$ 
of the coordinate system. The particles are either two spheres or two macroscopically long, parallel 
cylinders with axes in $y$-direction; their cross-sections in the plane $y=0$ are shown in 
Fig.~\ref{Fig1}. The fluid-fluid interface is indicated by the horizontal line at $x=0$. Although 
the particles can differ in size, they are taken to be chemically identical such that the surfaces 
of both particles in contact with the same fluid phase carry the same surface charge density, and 
that the contact angle $\theta$ is the same for both particles. This is a simplifying assumption 
because chemically identical particles in general need not to be equally charged \cite{Maj18}. In 
equilibrium the centers of the left and the right particle are located at $x=D_1$ 
and $x=D_2$, respectively. Depending upon the contact angle $\theta\in\left[0,\pi\right]$, both 
$D_1=-R_1\cos\theta$ and $D_2=-R_2\cos\theta$ can be negative (for $\theta<\frac{\pi}{2}$; the 
case considered in Fig.~\ref{Fig1}) as well as positive (for $\theta>\frac{\pi}{2}$). In between 
the particles a gap of width $L$ occurs (see the vertical dashed lines in Fig.~\ref{Fig1}) so that 
the horizontal center-to-center distance is $L+R_1+R_2$. The fluid phase below (above) the interface 
occupying the half-space $x<0~(x>0)$ is denoted by medium ``1'' (``2''). Both fluids are modeled 
as structureless, continuous media with dielectric constant $\varepsilon_i=\varepsilon_{r,i}\varepsilon_0$, 
$i\in\left\{1,2\right\}$, where $\varepsilon_{r,i}$ is the relative premittivity of medium $i$ and 
$\varepsilon_0$ is the vacuum permittivity. The ionic strength of added salt in medium 
$i\in\left\{1,2\right\}$ is denoted by $I_i$. The corresponding Debye screening length in each medium 
is given by $\kappa_i^{-1}=\sqrt{\varepsilon_{r,i}/\left(8\pi\ell_BI_i\right)}$ where 
$\ell_B=e^2/\left(4\pi\varepsilon_0k_BT\right)$ is the vacuum Bjerrum length with $e>0$, $k_B$, 
and $T$ being the elementary charge, the Boltzmann constant, and the absolute temperature, respectively. 
Within our description at the mean-field level, the length scale of interest is the Debye length 
as both the local charge density and the electrostatic interaction vary on this scale. Phenomena 
which occur on smaller length scales, e.g., the structuring of liquids on the molecular length 
scale, are not considered. Here we consider medium ``2'' to be the less polar phase in the sense 
that $\kappa_2^{-1}>\kappa_1^{-1}$. With this the criteria for the applicability of the DA are 
that both radii $R_1$ and $R_2$ have to be much larger than $\kappa_2^{-1}$ and $L$.

Within the DA, one basically decomposes the two interacting particles into infinitesimal 
surface elements. Assuming that the elementary surface pieces, which face each other, interact 
like flat parallel surfaces, the total interaction between the two curved objects is obtained 
via integration over the whole surface. Here, however, the particle surfaces are homogeneously 
charged only separately inside each medium, the properties of which in general differ. As a 
result, a three-phase contact line is formed where a particle surface intersects the fluid 
interface; two such contact lines on opposing particles interact as well. But this does 
not introduce any additional constraint for applying the DA. In the spirit of the DA, each 
of these two contact lines can be divided in infinitesimal pieces and the total contribution 
due to the line interaction can be obtained as long as the interaction between two parallel 
lines is known. Therefore, in order to apply the DA, one needs to know the interaction of 
parallel flat surfaces dipped into medium ``1'' or medium ``2'' and the interaction between 
two parallel three-phase contact lines. These are exactly the quantities we calculated in 
Ref.~\cite{Maj16} numerically by solving the nonlinear PB equation. To be more precise, the 
relevant quantities, as defined in Ref.~\cite{Maj16}, are $\omega_{\gamma,i}(r)$, which is 
the interaction energy per total surface area between two parallel, planar surfaces dipped 
at a distance $r$ into medium $i\in\{1,2\}$, and $\omega_\tau(r)$, which is the interaction 
energy per total line length between two parallel three-phase contact lines at a distance $r$. 
Please note that the interaction of the surface of one particle in contact with medium ``1'' 
and that of the other particle in contact with medium ``2'' is included in the line contribution 
$\omega_\tau(r)$. In order to tackle the problem efficiently, we 
first fit simple functions to the numerical data for $\omega_{\gamma,1}(r)$, $\omega_{\gamma,2}(r)$, 
and $\omega_{\tau}(r)$ obtained by full minimization of the nonlinear PB grand potential.
It turns out that a reasonably good fitting can be obtained by superposing exponential 
contributions as follows:
\begin{align}
 \omega_{\gamma,1}(r)=\sum\limits_{i=1}^3a_i\exp\left(-b_ir\right),
 \label{eq:1}
\end{align}
\begin{align}
 \omega_{\gamma,2}(r)=\sum\limits_{i=1}^3c_i\exp\left(-d_ir\right),
 \label{eq:2}
\end{align}
and
\begin{align}
 \omega_{\tau}(r)=\sum\limits_{i=1}^4g_i\exp\left(-h_ir\right).
 \label{eq:3}
\end{align}
For two flat plates, all interactions decay 
exponentially in the limit of large separations: $\omega_{\gamma,i}(r\to\infty)\sim\exp(-\kappa_ir)$ and 
$\omega_{\tau}(r\rightarrow\infty)\sim\exp(-\kappa_2r)$ (note the convention $\kappa_2^{-1}>\kappa_1^{-1}$). 
As a result, when fitting the data over a sufficiently large interval of $r$ (i.e., a 
few Debye lengths), the slowest of the decay rates $b_i$ in Eq.~(\ref{eq:1}) equals $\kappa_1$ 
and the slowest of the decay rates $d_i$ in Eq.~(\ref{eq:2}) as well as the slowest of the 
decay rates $h_i$ in Eq.~(\ref{eq:3}) equal $\kappa_2$. All three interactions, i.e., $\omega_{\gamma,1}$, 
$\omega_{\gamma,2}$, and $\omega_{\tau}$ result in forces onto the particles, which can be 
obtained by taking the negative derivative with respect to the appropriate distance between 
the facing surface or line elements, followed by integrating over the particles. Due to 
the geometry of the problem, the electrostatic force between the particles acts only 
in the horizontal $z$-direction. Please note that the movement of the particles in the 
vertical $x$-direction is suppressed by the steep and strong trapping potential \cite{Bin06}. 
In the following, we denote the $z$-component of the electrostatic force, which the left 
particle in Fig.~\ref{Fig1} exerts on the right one, by $F(L)$, and we decompose 
it, according to $F(L)=F_1(L)+F_2(L)+F_3(L)$, into the surface contribution $F_1(L)$ 
due to the surface interaction $\omega_{\gamma,1}$ in medium ``1'', the surface contribution 
$F_2(L)$ due to the surface interaction $\omega_{\gamma,2}$ in medium ``2'', and the line 
contribution $F_3(L)$ due to the line interaction $\omega_{\tau}$.

\section{Results and discussion}

In this section we discuss the variation of the force $F(L)$ between the particles as function of 
the particle sizes (radii $R_1$ and $R_2$), the particle separation $L$, and the contact angle 
$\theta$. For our discussion we consider two typical experimental setups and in each 
case the results for a pair of spheres as well as a pair of parallel cylinders are presented. Between the 
two systems considered below, the data for flat wall interactions for water-lutidine interfaces are 
taken from Ref.~\cite{Maj16} and those for water-octanol interfaces are newly generated here. We 
mention that all numerical examples presented here have been chosen such that the conditions for 
applying the DA are satisfied. Consequently, systems featuring oil with \textit{very} low 
dielectric constants, such as decane or octane, have been excluded because the corresponding 
Debye length $\kappa_2^{-1}$ is too large for them to satisfy the condition $R_1,R_2>>\kappa_2^{-1}$, even 
for micron size colloids. 

\subsection{Water-lutidine interface}
First, we consider a system consisting of polystyrene particles placed at a water-lutidine (2,6-dimethylpyridine) 
interface at temperature $T=313\,\mathrm{K}$. The added salt is NaI with bulk ionic strengths $I_1=1\,\mathrm{mM}$ 
and $I_2=0.85\,\mathrm{mM}$. The relative permittivities are $\varepsilon_{r,1}=72$ for the 
water-rich phase (medium ``1'') and $\varepsilon_{r,2}=62$ for the lutidine-rich phase 
(medium ``2''). The chemically identical particles are assumed to be similarly charged; the 
magnitude of the surface charge density in contact with the aqueous phase is $\sigma_1=0.1\,e/\mathrm{nm^2}$ 
and that in contact with the oil-phase is $\sigma_2=0.01\,e/\mathrm{nm^2}$. Differences in the 
solubilities of the ions in the two fluids result in a potential difference between the bulk of the 
two media, which is called the Donnan potential or Galvani potential difference \cite{Bag06}, and 
which, for our system, is assumed to be $1\,k_BT/e$. These numbers correspond to a standard set 
of parameters as used in Ref.~\cite{Maj16}. They are either taken or estimated from various 
experimental studies \cite{Ram58, Gal92, Gra93, Ine94, Lid98, Ave02, Bie12}. 

\subsubsection{Spheres}

\begin{figure*}[!t]
\centering{\includegraphics[width=16.0cm]{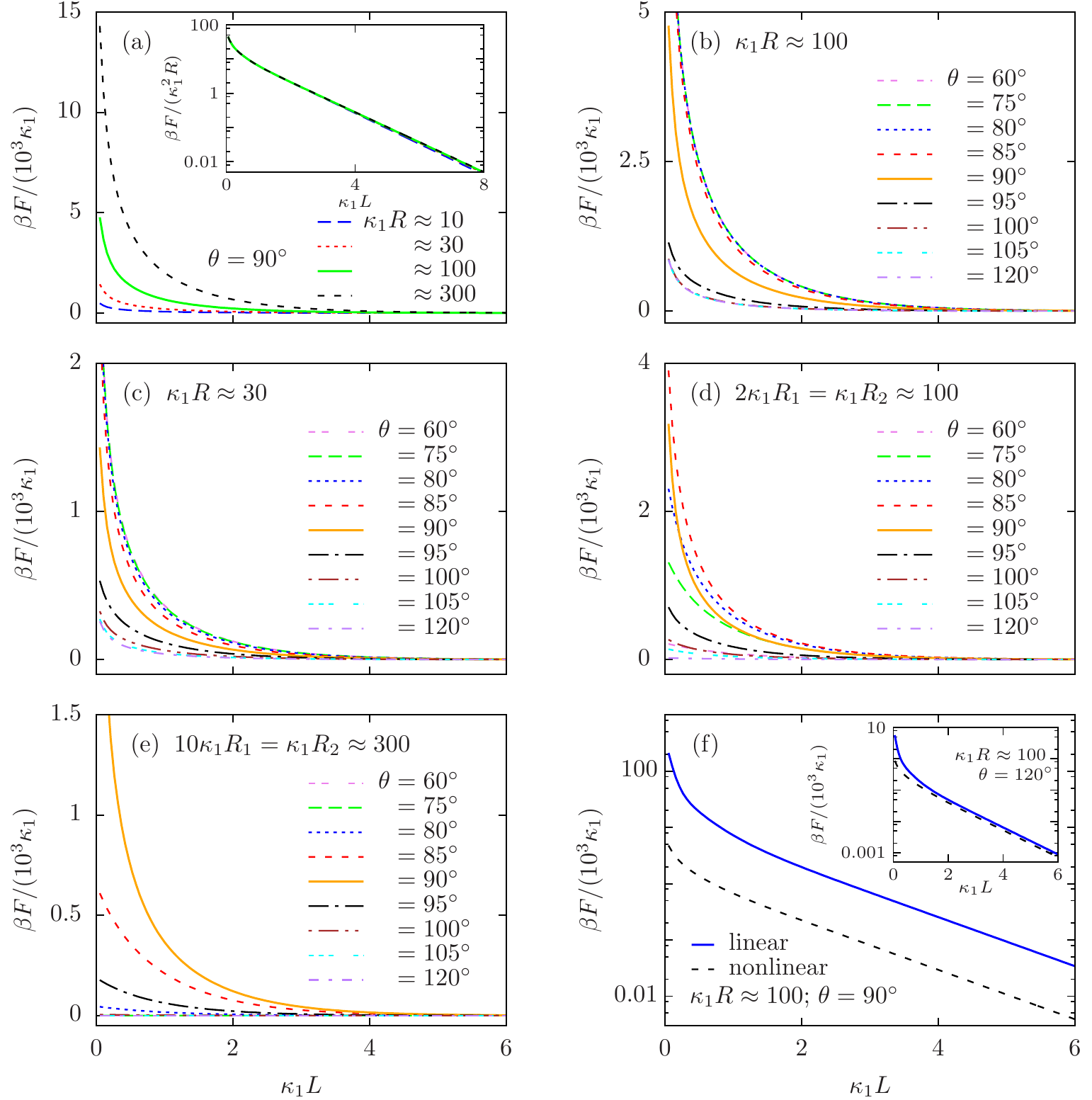}}
\caption{Variation of the lateral component of the force $F(L)$ due to the electrostatic 
         interaction between a pair of spherical colloidal particles, expressed in units 
         of $10^3\kappa_1/\beta$, as function of their scaled separation $\kappa_1L$ for 
         (a) equally sized ($R_1=R_2=R$) spheres of varying radius with contact angle $\theta=90^\circ$, 
         (b) equally sized ($\kappa_1R\approx100$) particles with varying $\theta$, 
         (c) equally sized ($\kappa_1R\approx30$) particles with varying $\theta$, 
         (d) unequally sized ($2\kappa_1R_1=\kappa_1R_2\approx100$) particles with varying $\theta$, 
         (e) unequally sized ($10\kappa_1R_1=\kappa_1R_2\approx300$) particles with varying $\theta$, and 
         (f) equally sized ($\kappa_1R\approx100$) particles with $\theta=90^\circ$ within 
         linear and nonlinear PB theory. As shown by panel (a) and its inset, the force 
         increases linearly with increasing $R$ and decays exponentially with increasing 
         separation $\kappa_1L$. Panels (b) and (c) suggest that, for equally sized 
         spheres, the force increases significantly with decreasing contact angle $\theta$ 
         only within an interval around $90^\circ$. Outside this interval the force 
         remains de facto constant and the interval of $\theta$, across which the force 
         actually varies, widens upon decreasing $\kappa_1R$. For unequally sized spheres, 
         if the size asymmetry is moderate, the force may increase as well as decrease if 
         the contact angle deviates from $90^\circ$ (panel (d)). However, if the size contrast 
         is high, the force becomes weaker once $\theta$ is slightly shifted away from $90^\circ$ 
         in either direction (panel (e)). From panel (f) and the inset therein one can infer 
         that the discrepancy between the linear and the nonlinear results diminishes 
         with increasing $\theta$.}
\label{Fig2}
\end{figure*}

In the case of two interacting spheres at a fluid interface, after performing the surface and 
line integrations, for the three distinct, lateral force contributions the following expressions 
are obtained:
\begin{align}
 F_1(L)=&\sum\limits_{i=1}^3\frac{\pi a_i}{\left(\frac{1}{R_1}+\frac{1}{R_2}\right)}
         \exp\left\{-b_i\left(L+\frac{\left(D_1-D_2\right)^2}{2\left(R_1+R_2\right)}\right)\right\}\times\notag\\
        &\left[1-\erf\left(\left(D_1R_2+D_2R_1\right)\sqrt{\frac{b_i}{2R_1R_2\left(R_1+R_2\right)}}~\right)\right],
 \label{eq:4}
\end{align}
\begin{align}
 F_2(L)=&\sum\limits_{i=1}^3\frac{\pi c_i}{\left(\frac{1}{R_1}+\frac{1}{R_2}\right)}
         \exp\left\{-d_i\left(L+\frac{\left(D_1-D_2\right)^2}{2\left(R_1+R_2\right)}\right)\right\}\times\notag\\
        &\left[1+\erf\left(\left(D_1R_2+D_2R_1\right)\sqrt{\frac{d_i}{2R_1R_2\left(R_1+R_2\right)}}~\right)\right],
 \label{eq:5}
\end{align}
where $\erf(x)$ denotes the error function \cite{Abr64}, and
\begin{align}
 F_3(L)=\sum\limits_{i=1}^4g_i\sqrt{\frac{2\pi h_i}{\left(\frac{1}{R_1}+\frac{1}{R_2}\right)}}
        \exp\left\{-h_i\left(L+\!\frac{D_1^2}{2R_1}\!+\!\frac{D_2^2}{2R_2}\right)\right\}
 \label{eq:6}
\end{align}
with $D_1=-R_1\cos\theta$ and $D_2=-R_2\cos\theta$. Variations of the total force $F(L)=F_1(L)+F_2(L)+F_3(L)$ 
in the units of $10^3\kappa_1/\beta$ with the scaled separation $\kappa_1L$ for different system parameters 
are shown in Fig.~\ref{Fig2}. Here $\beta=1/\left(k_BT\right)$ is the inverse thermal energy. As one 
can infer from Fig.~\ref{Fig2}(a), for equally-sized spheres ($R_1=R_2=R$) with a contact angle of 
$\theta=90^\circ$, the effective force scales linearly with the size of the particles and decays 
exponentially with increasing separation between the particles. Both the linear scaling with $R$ and 
the exponential decay with $L$ can be directly inferred from the inset of Fig.~\ref{Fig2}(a) where 
the ratio of the dimensionless force $\beta F/\kappa_1$ to the scaled radius $\kappa_1R$ is plotted as 
a function of the scaled separation $\kappa_1L$ using a semi-logarithmic scale, revealing data collapse.
The exponential decay is expected to occur in the sense that all effective interactions 
decay exponentially for a pair of interacting flat plates, which remains unaffected while using the DA, and 
the linear scaling with $R$ is a direct consequence of the DA.
Figures~\ref{Fig2}(b) and \ref{Fig2}(c) display the variation of the scaled force with contact angle $\theta$ 
for equally-sized particles. As one can see, for $\kappa_1R\approx100$ (b), the force increases with 
decreasing contact angle but de facto it varies only within the interval $80^\circ<\theta<100^\circ$. A 
similar phenomenon is observed for $\kappa_1R\approx30$ (c) albeit with variation in a slightly broader 
interval $75^\circ<\theta<105^\circ$. Figures~\ref{Fig2}(d) and \ref{Fig2}(e) show the variation of the 
force with contact angle $\theta$ as function of separation distance $\kappa_1L$ for unequally-sized 
spheres. For the relatively small size-asymmetry in Fig.~\ref{Fig2}(d), the force between the particles 
can increase in a certain range of separation between the particles as the contact angle $\theta$ is 
decreased slightly from $90^\circ$ downwards. Otherwise the force is weaker than the one obtained for 
a neutral wetting situation $\theta=90^\circ$. For the larger size-asymmetry in Fig.~\ref{Fig2}(e), the 
force becomes weaker as soon as the contact angle $\theta$ differs from $90^\circ$. However, in both 
cases, as considered in Figs.~\ref{Fig2}(d) and \ref{Fig2}(e), the force between the particles decreases 
upon increasing $\theta$ in the interval $\theta>90^\circ$. Finally, in Fig.~\ref{Fig2}(f), we compare 
the force obtained within the linearized and the nonlinear PB theory for equally sized spheres. As one 
can see from the plot, the forces differ by almost an order of magnitude even at large separations such 
as $\kappa_1L\approx6$ for $\theta=90^\circ$. It turns out that with increasing contact angle $\theta$ 
this difference diminishes; see the inset of Fig.~\ref{Fig2}(f), where both curves are of almost the 
same magnitude. This is expected since the portion of the particles immersed in the more polar phase, 
for which the electrostatic interaction is stronger because $\kappa_1\approx\kappa_2$ 
$\left[\kappa_1\approx0.1059\,\mathrm{nm^{-1}},\,\kappa_2\approx0.1053\,\mathrm{nm^{-1}}\right]$
and $\sigma_1\gg\sigma_2$, decreases with increasing contact angle. Note that this situation differs 
from the one in Refs. \cite{Hor03, Hor05} in which $\kappa_1\gg\kappa_2$. A similar discrepancy 
appeared while comparing the interactions within the linear and nonlinear theory for parallel flat 
surfaces \cite{Maj16}. Thus taking into account particle curvature does not significantly change this 
result.

All these observations are in accordance with the force expressions given in Eqs.~(\ref{eq:4})-(\ref{eq:6}). 
For micron-sized particles considered here, in most of the cases the line contribution to the total 
interaction is negligible. Therefore, the total force $F(L)$ is dominated by the surface 
contributions $F_1(L)$ (Eq.~(\ref{eq:4})) and $F_2(L)$ (Eq.~(\ref{eq:5})). For 
$\theta=90^\circ$, which is the case considered in Fig.~\ref{Fig2}(a), one has $D_1=D_2=0$. Moreover, 
with $R_1=R_2=R$, Eqs.~(\ref{eq:4}) and (\ref{eq:5}) reduce to 
$F_1(L)=\frac{\pi R}{2}\sum\limits_{i=1}^3a_i\exp\left(-b_iL\right)$ and 
$F_2(L)=\frac{\pi R}{2}\sum\limits_{i=1}^3c_i\exp\left(-d_iL\right)$, respectively, which 
transparently explain the linear variation of the force with the particle size $R$ (the coefficients 
$a_i$ and $c_i$ do not depend on $R$) and its exponential decay as function of the separation $L$. The 
decay rate in the limit of large distances is determined by the smaller of the two 
Debye lengths $\kappa_1^{-1}$ and $\kappa_2^{-1}$ (please note that for our system 
$\kappa_1^{-1}\approx\kappa_2^{-1}$). For $R_1=R_2=R$ with an arbitrary contact angle $\theta$, 
one has $D_1-D_2=0$ so that the dependence on $\theta$ appears in Eqs.~(\ref{eq:4}) and 
(\ref{eq:5}) only through the terms involving the error functions, which reduce to 
$1-\erf\left(-\cos\theta\sqrt{b_iR}\right)$ and $1+\erf\left(-\cos\theta\sqrt{d_iR}\right)$, respectively. 
Since the error function levels off to $|\erf(x)| \approx 1$ for $|x|\gtrsim2$, the 
variation of the force with respect to $\theta$ in Eq.~(\ref{eq:4}) saturates once the slowest 
decay rate $b_i$, which in the present case is $\kappa_1$, satisfies the inequality
\begin{align}
 |\cos\theta|\gtrsim\frac{2}{\sqrt{\kappa_1R}}.
 \label{eq:7}
\end{align}
Similarly, in Eq.~(\ref{eq:5}) the saturation is obtained once the slowest decay rate $d_i$, 
i.e., $\kappa_2$ satisfies the inequality
\begin{align}
 |\cos\theta|\gtrsim\frac{2}{\sqrt{\kappa_2R}}.
 \label{eq:7a}
\end{align}
Note that once the slowest decay rates $b_i$ and $d_i$ satisfy these conditions, all the 
other decay rates will do so, too. For $\kappa_1R\approx100$ (and therefore, $\kappa_2R\approx100$ 
as $\kappa_1\approx\kappa_2$), both Eqs.~(\ref{eq:7}) and (\ref{eq:7a}) predict that the force 
varies appreciably only within the interval $78^\circ\leq\theta\leq102^\circ$, which one precisely 
observes in Fig.~\ref{Fig2}(b). Decreasing the contact angle $\theta$ from 
$\frac{\pi}{2}$ implies that the particles become more hydrophilic. Consequently, the 
contribution $F_1(L)$ to the total force increases while $F_2(L)$ decreases upon 
decreasing $\theta$. Finally, at $\theta\approx78^\circ$ the former attains 
a non-zero finite value and the latter vanishes (please note the different signs in front 
of the error functions in Eqs.~(\ref{eq:4}) and (\ref{eq:5})). On the other hand, increasing 
the contact angle $\theta$ from $\frac{\pi}{2}$ implies that the particles 
become more hydrophobic. As a result, the contribution $F_2(L)$ increases and $F_1(L)$ 
decreases with the former attaining a non-zero finite value while the latter 
is vanishing at $\theta\approx102^\circ$. Upon decreasing $\kappa_1R$, the interval of $\theta$ 
over which the force varies broadens as can be inferred from Fig.~\ref{Fig2}(c). For unequal 
particles sizes, i.e., for $R_1\neq R_2$, the dependence on $\theta$ in Eqs.~(\ref{eq:4}) and 
(\ref{eq:5}) originate from both the exponential and the error function. For moderate size-asymmetry, 
like the one considered in Fig.~\ref{Fig2}(d), a competition between these two functions determines 
the variation of the force with $\theta$. However, for the extremely asymmetric case considered 
in Fig.~\ref{Fig2}(e), the difference $D_1-D_2$ is large and the exponential terms dominate 
as soon as $\theta$ differs slightly from $90^\circ$.

\begin{figure*}[!t]
\centering{\includegraphics[width=16.0cm]{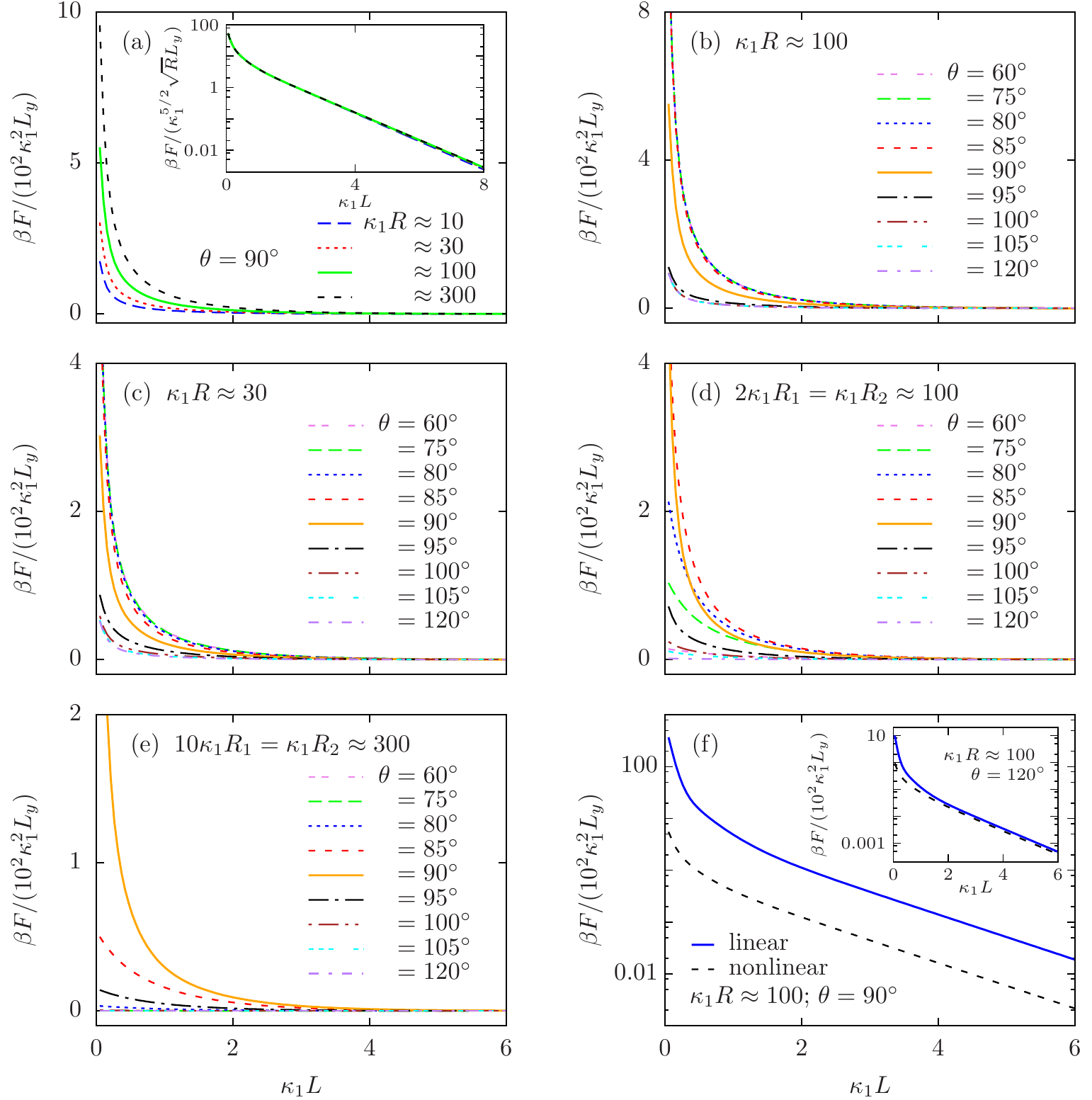}}
\caption{Variation of the $z$-component (see Fig.~\ref{Fig1}) of the force $F(L)$, per length $L_y$ 
         in the $y$-direction, due to the electrostatic interaction between a pair of 
         parallel cylindrical colloids, expressed in units of $10^2\kappa_1^2/\beta$, as 
         function of their scaled separation $\kappa_1L$ for 
         (a) equally sized ($R_1=R_2=R$) cylinders of varying radius with contact angle $\theta=90^\circ$, 
         (b) equally sized ($\kappa_1R\approx100$) particles with varying $\theta$, 
         (c) equally sized ($\kappa_1R\approx30$) particles with varying $\theta$, 
         (d) unequally sized ($2\kappa_1R_1=\kappa_1R_2\approx100$) particles with varying $\theta$,
         (e) unequally sized ($10\kappa_1R_1=\kappa_1R_2\approx300$) particles with varying $\theta$, and 
         (f) equally sized ($\kappa_1R\approx100$) particles with $\theta=90^\circ$ within 
         linear and nonlinear PB theory. As shown by panel (a) and its inset, the force 
         increases $\propto\sqrt{R}$ with increasing $R$ and decays exponentially with 
         increasing separation $\kappa_1L$. Panels (b) and (c) suggest that, for 
         equally sized cylinders, the force increases significantly with decreasing
         contact angle $\theta$ only within an interval around $90^\circ$. Outside this 
         interval the force remains de facto constant and the interval of $\theta$, across 
         which the force actually varies, widens upon decreasing $\kappa_1R$. For unequally sized 
         cylinders, if the size asymmetry is moderate, the force may increase as well as 
         decrease if the contact angle deviates from $90^\circ$ (panel (d)). However, if the size 
         contrast is high, the force becomes weaker once $\theta$ is slightly shifted away from 
         $90^\circ$ in either direction (panel (e)). From panel (f) and the inset therein 
         one can infer that the discrepancy between the linear and the nonlinear results 
         diminishes with increasing $\theta$.}
\label{Fig3}
\end{figure*}

\subsubsection{Cylinders}

For macroscopically long, parallel cylinders, the expressions for the lateral force 
contributions, expressed per length $L_y$ in the $y$-direction, are given by:
\begin{align}
 \frac{F_1(L)}{L_y}&=\sum\limits_{i=1}^3a_i\sqrt{\frac{\pi b_i}{\left(\frac{2}{R_1}+\frac{2}{R_2}\right)}}
         \exp\left\{-b_i\left(L+\frac{\left(D_1-D_2\right)^2}{2\left(R_1+R_2\right)}\right)\right\}\times\notag\\
        &\left[1-\erf\left(\left(D_1R_2+D_2R_1\right)\sqrt{\frac{b_i}{2R_1R_2\left(R_1+R_2\right)}}~\right)\right],
 \label{eq:8}
\end{align}
\begin{align}
 \frac{F_2(L)}{L_y}&=\sum\limits_{i=1}^3c_i\sqrt{\frac{\pi d_i}{\left(\frac{2}{R_1}+\frac{2}{R_2}\right)}}
         \exp\left\{-d_i\left(L+\frac{\left(D_1-D_2\right)^2}{2\left(R_1+R_2\right)}\right)\right\}\times\notag\\
        &\left[1+\erf\left(\left(D_1R_2+D_2R_1\right)\sqrt{\frac{d_i}{2R_1R_2\left(R_1+R_2\right)}}~\right)\right],
 \label{eq:9}
\end{align}
and
\begin{align}
 \frac{F_3(L)}{L_y}=\sum\limits_{i=1}^4g_ih_i\exp\left\{-h_i\left(L+\!\frac{D_1^2}{2R_1}\!+\!\frac{D_2^2}{2R_2}\right)\right\}.
 \label{eq:10}
\end{align}
We note that for geometrical reasons these expressions are slightly different from those obtained 
for spheres in Eqs.~(\ref{eq:4})--(\ref{eq:6}). In particular, the contact lines for cylinders are 
just straight lines and in order to obtain $F_3(L)$ there is no need to use the DA. Figure~\ref{Fig3} 
shows the variation of the $z$-component of the total force $F(L)=F_1(L)+F_2(L)+F_3(L)$, per length 
$L_y$ in the $y$-direction and in units of $10^2\kappa_1^2/\beta$, which the left cylinder 
exerts on the right one as function of the scaled separation $\kappa_1L$ for the sizes $R_1$ and 
$R_2$ and for the contact angle $\theta$. Except for a few features, the findings are qualitatively 
the same as those obtained for spheres in Fig.~\ref{Fig2}. For example, for $R_1=R_2=R$ and 
$\theta=90^\circ$, the force between two cylinders also decays exponentially with varying separation 
$L$ between them and increases with increasing size $R$, but for the cylinders the increase is 
proportional to $\sqrt{R}$; see Fig.~\ref{Fig3}(a) and the inset therein. This is evident from the 
prefactors of the exponential functions in Eqs.~(\ref{eq:8}) and (\ref{eq:9}). The variation of the 
force with respect to $\theta$ as well as Eqs.~(\ref{eq:7}) and (\ref{eq:7a}) remain the same as 
for the spheres, because the $\theta$-dependent terms in Eqs.~(\ref{eq:8}) and (\ref{eq:9}) have 
exactly the same form as in Eqs.~(\ref{eq:4}) and (\ref{eq:5}). This behavior is confirmed by 
Figs.~\ref{Fig3}(b)--\ref{Fig3}(e). Finally, the comparison of the effective force for $\theta=90^\circ$ 
within linearized and nonlinear PB theory reveals a significant discrepancy between the predictions 
of the two approaches, which becomes smaller for larger contact angles $\theta$, i.e., as the 
portion of the particles, dipped into the more polar phase, decreases (see Fig.~\ref{Fig3}(f)).

\begin{figure*}[!t]
\centering{\includegraphics[width=16.2cm]{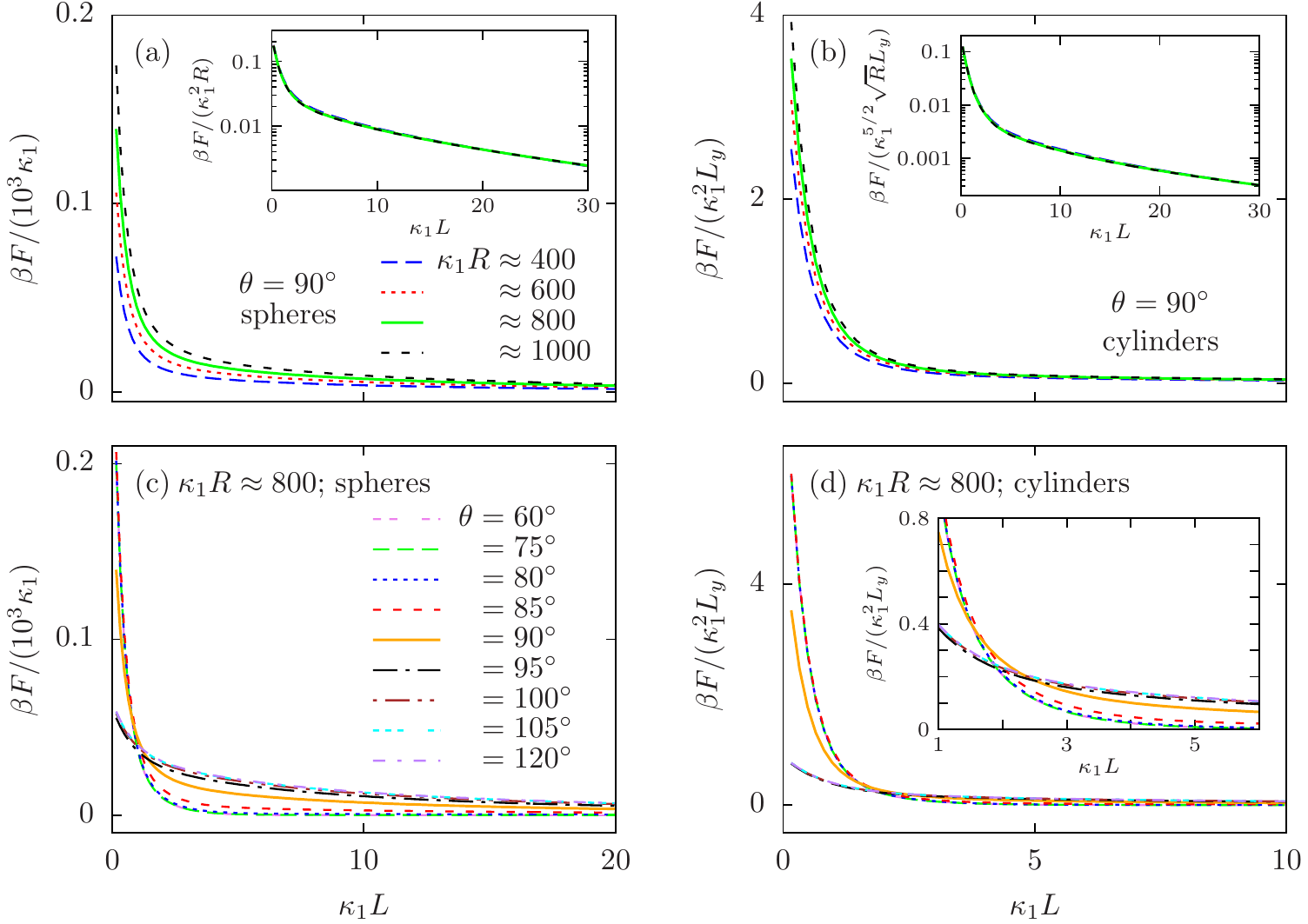}}
\caption{Panel (a): Variation of the lateral component of the force $F(L)$, 
         expressed in units of $10^3\kappa_1/\beta$, due to electrostatic interaction between a 
         pair of equally sized $(R_1=R_2=R)$ spheres of varying radius with contact angle $90^\circ$ as 
         function of their scaled separation $\kappa_1L$. The force increases linearly with 
         increasing size of the particles which is evident from the data collapse in the inset. 
         Panel (b): Variation of the $z$-component (see Fig.~\ref{Fig1}) of the force $F(L)$, 
         per length $L_y$ in the $y$-direction, expressed in units of $\kappa_1^2/\beta$, 
         due to electrostatic interaction between a pair of equally sized cylinders of varying 
         radius with contact angle $90^\circ$, as function of their scaled separation $\kappa_1L$. 
         Contrary to what one observes for spheres, the force between cylinders is proportional to $\sqrt{R}$.
         Panel (c): Variation of the lateral component of the force $F(L)$, 
         expressed in units of $10^3\kappa_1/\beta$, due to electrostatic interaction between a 
         pair of equally sized $(\kappa_1R\approx800)$ spheres with various 
         contact angles $\theta$, as function of their scaled separation $\kappa_1L$. 
         Panel (d): Variation of the $z$-component of the force $F(L)$, per length $L_y$ in 
         the $y$-direction, expressed in units of $\kappa_1^2/\beta$, due to electrostatic 
         interaction between a pair of equally sized $(\kappa_1R\approx800)$ cylinders for 
         various contact angles $\theta$, as function of their scaled separation $\kappa_1L$. 
         Both for spheres and cylinders the force varies within a narrow interval of the 
         contact angle at very short separations. The force increases if the particles are 
         more hydrophilic within this interval of $\theta$. 
         At relatively large separations, however, this interval slightly broadens but the force 
         increases if the particles become more hydrophobic.}
\label{Fig4}
\end{figure*}

\subsection{Water-octanol interface}

Water and lutidine, which are immiscible for sufficiently high temperatures, form a 
special system in that the bulk properties, i.e., the relative permittivities and the bulk ionic 
strengths, and consequently the Debye screening lengths, are not very different for the two fluid 
phases. In contrast to that, in the present subsection we consider another system with silica 
particles trapped at a water-octanol interface. At room temperature $T=300\,\mathrm{K}$ these two 
fluids differ starkly with respect to their bulk properties with $\varepsilon_{r,1}=80$ 
for water and $\varepsilon_{r,2}=10.3$ for octanol. The partitioning of ions at such an interface leads 
to highly contrasting bulk ionic strengths: for $I_1=10\,\mathrm{mM}$ one 
has $I_2=2.9\times10^{-3}\,\mathrm{mM}$; the corresponding resulting Donnan potential 
equals $3.8\,k_BT/e$ \cite{Gul02, Que08}. Under these conditions, the inverse Debye length in 
the water phase is $\kappa_1\approx0.324\,\mathrm{nm^{-1}}$ and the one in the oil phase (octanol) 
is $\kappa_2\approx0.015\,\mathrm{nm^{-1}}$. The magnitude of the surface charge densities 
in contact with the two fluid phases also differ significantly; we consider 
$\sigma_1=0.01\,e/\mathrm{nm^2}$ and $\sigma_2=0.0005\,e/\mathrm{nm^2}$ \cite{Dan04, Hor03, Hor05}. 

The resulting interactions between the particles are shown in Fig.~\ref{Fig4} 
for a pair of spheres (panels (a) and (c)) as well as for a pair of cylinders (panels (b) and (d)). 
From Figs.~\ref{Fig4}(a) and (b) one can infer that the total force $F(L)$ between equally sized 
particles increases with increasing radii $(R_1=R_2=R)$, both for spheres and cylinders. Whereas for 
spheres this increase is linear in the particle size (see the inset in Fig.~\ref{Fig4}(a)), in 
the case of cylinders it scales $\propto\sqrt{R}$ (see the inset in Fig.~\ref{Fig4}(b)), which 
is evident from the data collapse in the insets. Although the line interaction becomes relatively 
more important in the case of the water-octanol system -- due to a greater mismatch of the system 
parameters (ionic strengths, permittivities, and charge densities) compared to those of the 
water-lutidine system -- these findings suggest that for micron-sized particles the interaction 
is still dominated by the surface parts. Figures~\ref{Fig4}(c) and (d) show the variation of the 
inter-particle forces $F(L)$ as function of the wetting contact angle $\theta$ for spheres and cylinders, respectively.
At very short separations, the force varies only within a narrow interval 
$85^\circ\lesssim\theta\lesssim95^\circ$. However, at relatively large separations it varies within 
a wider interval $75^\circ\lesssim\theta\lesssim105^\circ$ of the contact angle. These findings are also in 
accordance with Eqs.~(\ref{eq:7}) and (\ref{eq:7a}). For the system considered here, Eq.~(\ref{eq:7}) 
predicts that $F_1(L)$ varies appreciably within the interval $86^\circ\lesssim\theta\lesssim94^\circ$ 
whereas, according to Eq.~(\ref{eq:7a}), $F_2(L)$ varies within the interval 
$71^\circ\lesssim\theta\lesssim109^\circ$. At very short separations, the total force is dominated by the 
surfce contribution in medium ``1'' (aqueous phase) due to higher surface charge densities at the 
particle surfaces. Therefore, if $\theta$ is decreased from $90^\circ$, i.e., when the particles become 
increasingly hydrophilic, the force increases, followed by saturation at around 
$\theta=85^\circ$, as predicted by 
Eq.~(\ref{eq:7}). On the other hand, if $\theta$ is increased 
beyond $90^\circ$ the particles become more 
hydrophobic. Up to $\theta\approx95^\circ$, for which $F_1$ vanishes, 
the total force decreases as $F_1$ decreases. Beyond that, a slight increase of the total force 
is observed due to $F_2$ which, as predicted by Eq.~\ref{eq:7a}, increases up 
to $\theta\approx109^\circ$. As the separation between the 
particles is increased, in medium ``1'' the interaction decays very fast due to 
a strong screening by the 
higher amount of salt present. Consequently, at relatively large separations the total 
force is dominated by the surface contribution in medium ``2'' (oil phase) and, within 
the interval $75^\circ\lesssim\theta\lesssim105^\circ$, it increases monotonically with increasing 
contact angle. It is important to note that Eqs.~(\ref{eq:7}) and (\ref{eq:7a}) are 
derived by using the fact that the error function $\erf(x)$ saturates for 
$|x|\gtrsim2$, with the most significant variation occuring only 
for $|x|\lesssim1.5$. Therefore, the variation of $F_2(L)$ within the intervals 
$71^\circ\lesssim\theta\lesssim75^\circ$ and 
$105^\circ\lesssim\theta\lesssim109^\circ$ are very slow and hardly visible. Since here the 
silica particles are considered to be weakly charged, the discrepancy 
between the linear and the nonlinear PB theories become 
less significant. Still, the forces within the two approaches differ 
by a factor of $2$ even at separations $\kappa_1L\approx10$ for $\theta=90^\circ$. 

\section{Conclusions}

To conclude, by using the Derjaguin approximation and a fitting procedure for numerical
results for the effective interaction between parallel, planar surfaces in contact with 
two demixed fluids in between, we have calculated the force due to the electrostatic 
interaction between pairs of spheres or cylinders at close distance from each other at a 
fluid-fluid interface. The comparison between the results obtained within linear and 
nonlinear PB theory shows that the former overestimates the force both for spheres and 
for cylinders, even at distances of several Debye lengths. Concerning the results within 
the nonlinear theory, we have investigated the effects of varying the sizes and the 
contact angle of the particles. Our general study is applicable also to pairs of particles 
which differ in size. For equally-sized spheres and cylinders the force always decays 
exponentially with increasing separation, and it scales $\propto R$ for spheres and 
$\propto\sqrt{R}$ for cylinders, where $R$ is the common radius of the particles. Importantly, 
for equally-sized particles (both spherical and cylindrical) we have found an interval around 
the contact angle of $90^\circ$, beyond which the force de facto does not vary. We have 
also obtained simple relations (Eqs.~(\ref{eq:7}) and (\ref{eq:7a})) involving the Debye 
lengths of the two media and the radii of the particles for calculating the width of this 
interval. These robust results can be expected to be useful for describing more general 
or complex particle interactions at fluid interfaces, which is important for various 
application perspectives of such systems.



\begin{thebibliography}{00}
   \bibitem{Ram03} W.\ Ramsden, Proc.\ R.\ Soc.\ London \textbf{72}, 156 (1903).
   \bibitem{Bin06} B.\ P.\ Binks and T.\ S.\ Horozov, \textit{Colloidal Particles at Liquid Interfaces} 
                   (Cambridge University Press, Cambridge, 2006).
   \bibitem{Pic07} S.\ U.\ Pickering, J. Chem. Soc. Trans.\ \textbf{91}, 2001 (1907).
   \bibitem{Din02} A.\ D.\ Dinsmore, M.\ F.\ Hsu, M.\ G.\ Nikolaides, M.\ M\'arquez, A.\ R.\ Bausch, and D.\ A.\ Weitz, 
                   Science \textbf{298}, 1006 (2002).
   \bibitem{Li13} M.\ Li, R.\ L.\ Harbron, J.\ V.\ M.\ Weaver, B.\ P.\ Binks, and S.\ Mann, Nat.\ Chem.\ \textbf{5}, 
                  529 (2013).
   \bibitem{Bok07} A.\ B\"oker, J.\ He, T.\ Emrick, and T.\ P.\ Russell, Soft Matter \textbf{3}, 1231 (2007).
   \bibitem{Rey16} B.\ M.\ Rey, R.\ Elnathan, R.\ Ditcovski, K.\ Geisel, M.\ Zanini, M.\ A.\ Fernandez-Rodriguez, 
                   V.\ V.\ Naik, A.\ Frutiger, W.\ Richtering, T.\ Ellenbogen, N.\ H.\ Voelcker, and L.\ Isa, 
                   Nano Lett. \textbf{16}, 157 (2016).
   \bibitem{You05} T.\ Young, Philos.\ Trans.\ R.\ Soc.\ London \textbf{95}, 65 (1805).
   \bibitem{Gen85} P.\ G.\ de Gennes, Rev.\ Mod.\ Phys.\ \textbf{57}, 827 (1985).
   \bibitem{Ave03} R.\ Aveyard, B.\ P.\ Binks, and J.\ H.\ Clint, Adv.\ Colloid Interface Sci.\ \textbf{100}, 503 (2003).
   \bibitem{Hor03} T.\ S.\ Horozov, R.\ Aveyard, J.\ H.\ Clint, and B.\ P.\ Binks, Langmuir \textbf{19}, 2822 (2003).
   \bibitem{Hor05} T.\ S.\ Horozov, R.\ Aveyard, B.\ P.\ Binks, and J.\ H.\ Clint, Langmuir \textbf{21}, 7405 (2005).
   \bibitem{Sta00} D.\ Stamou, C.\ Duschl, and D.\ Johannsmann, Phys.\ Rev.\ E \textbf{62}, 5263 (2000).
   \bibitem{Kra00} P.\ A.\ Kralchevsky and K.\ Nagayama, Adv.\ Colloid Interface Sci.\ \textbf{85}, 145 (2000).
   \bibitem{Oet05} M.\ Oettel, A.\ Dom\'inguez, and S.\ Dietrich, Phys.\ Rev.\ E \textbf{71}, 051401 (2005).
   \bibitem{Lou05} J.\ C.\ Loudet, A.\ M.\ Alsayed, J.\ Zhang, and A.\ G.\ Yodh, Phys.\ Rev.\ Lett.\ \textbf{94}, 018301 
                   (2005).
   \bibitem{Oet08} M.\ Oettel and S.\ Dietrich, Langmuir \textbf{24}, 1425 (2008).
   \bibitem{Par15} L.\ Parolini, A.\ D.\ Law, A.\ Maestro, D.\ M.\ A.\ Buzza, and P.\ Cicuta, J.\ Phys.: Condens.\ Matter 
                   \textbf{27}, 194119 (2015).
   \bibitem{Ana16} S.\ E.\ Anachkov, I.\ Lesov, M.\ Zanini, P.\ A.\ Kralchevsky, N.\ A.\ Denkov, and L.\ Isa, Soft Matter
                   \textbf{12}, 7632 (2016).
   \bibitem{Pie80} P.\ Pieranski, Phys.\ Rev.\ Lett.\ \textbf{45}, 569 (1980).
   \bibitem{Hur85} A.\ J.\ Hurd, J. Phys. A \textbf{18}, L1055 (1985).
   \bibitem{Maj14} A.\ Majee, M.\ Bier, and S.\ Dietrich, J.\ Chem.\ Phys.\ \textbf{140}, 164906 (2014).
   \bibitem{Lia16} Z.\ Lian, J.\ Chem.\ Phys.\ \textbf{145}, 014901 (2016).
   \bibitem{Sch18} A.\ Majee, T.\ Schmetzer, and M.\ Bier, Phys.\ Rev.\ E \textbf{97}, 042611 (2018).
   \bibitem{Maj16} A.\ Majee, M.\ Bier, and S.\ Dietrich, J.\ Chem.\ Phys.\ \textbf{145}, 064707 (2016).
   \bibitem{Boo15} S.\ G.\ Booth and R.\ A.\ W.\ Dryfe, J.\ Phys.\ Chem.\ C \textbf{119}, 23295 (2015).
   \bibitem{Rus89} W.\ B.\ Russell, D.\ A.\ Saville, and W.\ R.\ Schowalter, \textit{Colloidal Dispersions} 
                   (Cambridge University Press, Cambridge, 1989).
   \bibitem{Lyn92} M.\ P.\ Lyne, B.\ D.\ Bowen, and S.\ Levine, J.\ Colloid Interface Sci.\ \textbf{150}, 374 (1992).
   \bibitem{Lop00} F.\ Mart\'inez-L\'opez, M.\ A.\ Cabrerizo-V\'ilchez, and R.\ Hidalgo-\'Alvarez, J.\ Colloid 
                   Interface Sci.\ \textbf{232}, 303 (2000).
   \bibitem{Ave00} R.\ Aveyard, J.\ H.\ Clint, D.\ Nees, and V.\ N.\ Paunov, Langmuir \textbf{16}, 1969 (2000).
   \bibitem{Ave02} R.\ Aveyard, B.\ P.\ Binks, J.\ H.\ Clint, P.\ D.\ I.\ Fletcher, T.\ S.\ Horozov, 
                   B.\ Neumann, V.\ N.\ Paunov, J.\ Annesley, S.\ W.\ Botchway, D.\ Nees, A.\ W.\ Parker, 
                   A.\ D.\ Ward, and A.\ N.\ Burgess, Phys.\ Rev.\ Lett.\ \textbf{88}, 246102 (2002).
   \bibitem{Par08} B.\ J.\ Park, J.\ P.\ Pantina, E.\ M.\ Furst, M.\ Oettel, S.\ Reynaert, and J.\ Vermant, 
                   Langmuir \textbf{24}, 1686 (2008).
   \bibitem{Gao14} P.\ Gao, X.\ C.\ Xing, Y.\ Li, T.\ Ngai, and F.\ Jin, Sci.\ Rep.\ \textbf{4}, 4778 (2014).
   \bibitem{Kel15} C.\ P.\ Kelleher, A.\ Wang, G.\ I.\ Guerrero-Garc\'ia, A.\ D.\ Hollingsworth, R.\ E.\ Guerra, 
                   B.\ J.\ Krishnatreya, D.\ G.\ Grier, V.\ N.\ Manoharan, and P.\ M.\ Chaikin, 
                   Phys.\ Rev.\ E \textbf{92}, 062306 (2015).
   \bibitem{Maj18} A.\ Majee, M.\ Bier, and R.\ Podgornik, Soft Matter \textbf{14}, 985 (2018).
   \bibitem{Bag06} V.\ S.\ Bagotsky, \textit{Fundamentals of Electrochemistry} (Wiley, Hoboken, NJ, 2006).
   \bibitem{Ram58} R.\ W.\ Rampolla and C.\ P.\ Smyth, J.\ Am.\ Chem.\ Soc.\ \textbf{80}, 1057 (1958).
   \bibitem{Gal92} P.\ D.\ Gallagher, M.\ L.\ Kurnaz, and J.\ V.\ Maher, Phys.\ Rev.\ A \textbf{46}, 7750 (1992).
   \bibitem{Gra93} C.\ A.\ Grattoni, R.\ A.\ Dawe, C.\ Y.\ Seah, and J.\ D.\ Gray, J.\ Chem.\ Eng.\ Data \textbf{38}, 
                   516 (1993).
   \bibitem{Ine94} H.\ D.\ Inerowicz, W.\ Li, and I.\ Persson, J.\ Chem.\ Soc.\ Faraday Trans.\ \textbf{90}, 2223 (1994).
   \bibitem{Lid98} D.\ R.\ Lide, \textit{Handbook of Chemistry and Physics, 82nd ed.} (CRC, Boca Raton, 2001$-$2002).
   \bibitem{Bie12} M.\ Bier, A.\ Gambassi, and S.\ Dietrich, J.\ Chem.\ Phys.\ \textbf{137}, 034504 (2012).
   \bibitem{Gul02} R.\ Gulaboski, V.\ Mir\v{c}eski, and F.\ Scholz, Electrochem.\ Commun.\ \textbf{4}, 277 (2002).
   \bibitem{Que08} F.\ Quentel, V.\ Mir\v{c}eski, and M.\ L'Her, J.\ Solid State Electrochem.\ \textbf{12}, 31 (2008).
   \bibitem{Dan04} K.\ D.\ Danov, P.\ A.\ Kralchevsky, and M.\ P.\ Boneva, Langmuir \textbf{20}, 6139 (2004).
   \bibitem{Abr64} M.\ Abramowitz and I.\ A.\ Stegun (eds.), \textit{Handbook of Mathematical Functions} (Dover, 
                   New York, 1964).
\end{thebibliography}
\end{document}